\documentclass[10pt,conference] {IEEEtran}
\IEEEoverridecommandlockouts

\usepackage{cite}
\usepackage{amsmath,amssymb,amsfonts}
\usepackage{graphicx}
\usepackage{textcomp}
\usepackage[table]{xcolor}
\usepackage{booktabs}
\usepackage{multirow}
\usepackage{subcaption}
\usepackage{balance}
\usepackage{url}
\usepackage{hyperref}
\hypersetup{hidelinks}
\usepackage{pifont}
\newcommand{\cmark}{\ding{51}}
\newcommand{\xmark}{\ding{55}}

\usepackage{tabularx}
\newcolumntype{Y}{>{\raggedright\arraybackslash}X}
\usepackage[most]{tcolorbox}
\tcbset{
  findingstyle/.style={
    enhanced, breakable,
    colback=black!5,
    colframe=black,
    boxrule=1.1pt,
    arc=1mm,
    left=2pt,right=2pt,top=1pt,bottom=1pt,
  }
}
\newtcolorbox{findingbox}[1][]{findingstyle,#1}

\DeclareRobustCommand{\name}{\textit{AgentTether}}
\newcommand{\taubench}{$\tau$-bench}
\newcommand{\pp}{\,pp}

\begin{document}

\title{\name{}: Graph-Guided Diagnosis and Runtime Intervention for Reliable LLM Agent Operations}


\author{
\IEEEauthorblockN{
Chenyu Zhao\textsuperscript{1},
Shenglin Zhang\textsuperscript{1,*},
Wenwei Gu\textsuperscript{1},
Yongqian Sun\textsuperscript{1}
}
\IEEEauthorblockN{
Dan Pei\textsuperscript{2},
Chetan Bansal\textsuperscript{3},
Saravan Rajmohan\textsuperscript{3},
and Minghua Ma\textsuperscript{3}
}

\IEEEauthorblockA{
\textsuperscript{1}~Nankai University, Tianjin, China
}

\IEEEauthorblockA{
\textsuperscript{2}~Tsinghua University, Beijing, China
}

\IEEEauthorblockA{
\textsuperscript{3}~Microsoft
}

\IEEEauthorblockA{
\textsuperscript{*}Corresponding author: Shenglin Zhang.
}
}

\maketitle

\begin{abstract}
Large language model (LLM) agents are increasingly used for multi-step, stateful tool-use tasks, yet production reliability remains limited.
Unlike static software repair, agent repair must recover dynamic trajectories whose early decisions can propagate into later errors and external state changes.
Existing automatic remedies address only part of this problem: blind retry adds no diagnosis, outcome feedback says \emph{whether} a run failed but not \emph{where} or \emph{why}, and self-reflection often lacks grounded evidence to prevent the same failure from recurring.
We present \name{}, a run-time repair framework that automates post-run diagnosis and guided recovery without modifying the underlying agent or environment.
\name{} abstracts each run into Transition Units, links them through a dependency-aware Critical Transition Graph, and localizes failure-critical subtrajectories by combining an offline normal-behavior model with a run-local graph detector.
It then converts the localized cause into behavior-scoped guidance backed by cross-iteration Repair Memory, and can optionally apply guarded run-time intervention to keep the correction active during re-execution.
The same design can be deployed as an offline diagnostic-and-guidance tool or as an online repair layer.

We evaluate \name{} on 261 \taubench{} tasks across three domains with Qwen3.7-max, and test cross-model transfer on Banking with GPT-5.4.
On the hardest Banking domain, \name{} repairs 59.04\% (49/83) of initially failed Qwen3.7-max tasks and 65.12\% (56/86) of initially failed GPT-5.4 tasks.
Overall, \name{} improves repair effectiveness while reducing agent turns and end-to-end approach tokens, suggesting a practical reliability layer that can wrap existing agent deployments, reduce wasted re-execution, and improve recovery without retraining the agent.
\end{abstract}


\section{Introduction}
\label{sec:intro}
Large language model (LLM) agents now execute multi-step tasks that require planning, tool use, state updates, and iterative interaction with external environments~\cite{xi2023rise, wang2024survey, yue2024tau, yao2023react, schick2023toolformer, yao2023tree, wang2023voyager}.
As they move from demonstrations to production, reliability becomes the central barrier: single-run success rates remain limited in realistic domains~\cite{yue2024tau, zhou2024webarena, liu2023agentbench, jimenez2024swebench, qin2024toolllm, patil2024gorilla, yang2024swe, kapoor2024agents}, and even seemingly successful runs may harbor process-level risks such as missing steps, incorrect ordering, tool-protocol violations, unauthorized state changes, or unnecessary actions~\cite{cemri2025mast, ruan2024identifying}.
Such defects may be missed by coarse outcome-based evaluation yet impose real operational and compliance costs, as illustrated by deployed chatbots that misstated refund rules or advised legally invalid actions~\cite{aircanada2024chatbot,nyc2024mycity}.
Improving reliability therefore demands moving beyond final-outcome measurement and blind retry, toward actively recovering the runs that go wrong.

We refer to this corrective process as \emph{agent repair}: diagnosing and correcting defective agent runs after failure or during execution, without modifying the underlying model or environment.
Unlike automated program repair or LLM-based self-debugging, which operate on fixed code artifacts and test oracles~\cite{legoues2019apr,xia2023apr,chen2024selfdebug}, agent repair targets dynamic, stateful, and non-deterministic trajectories whose failures may lack an oracle, originate far upstream, and change external state in ways that are hard to undo.
These differences make agent repair more than patch generation and motivate the expert human operator's repair loop as a reference.

\begin{figure*}[t]
\centering
\includegraphics[width=0.8\textwidth]{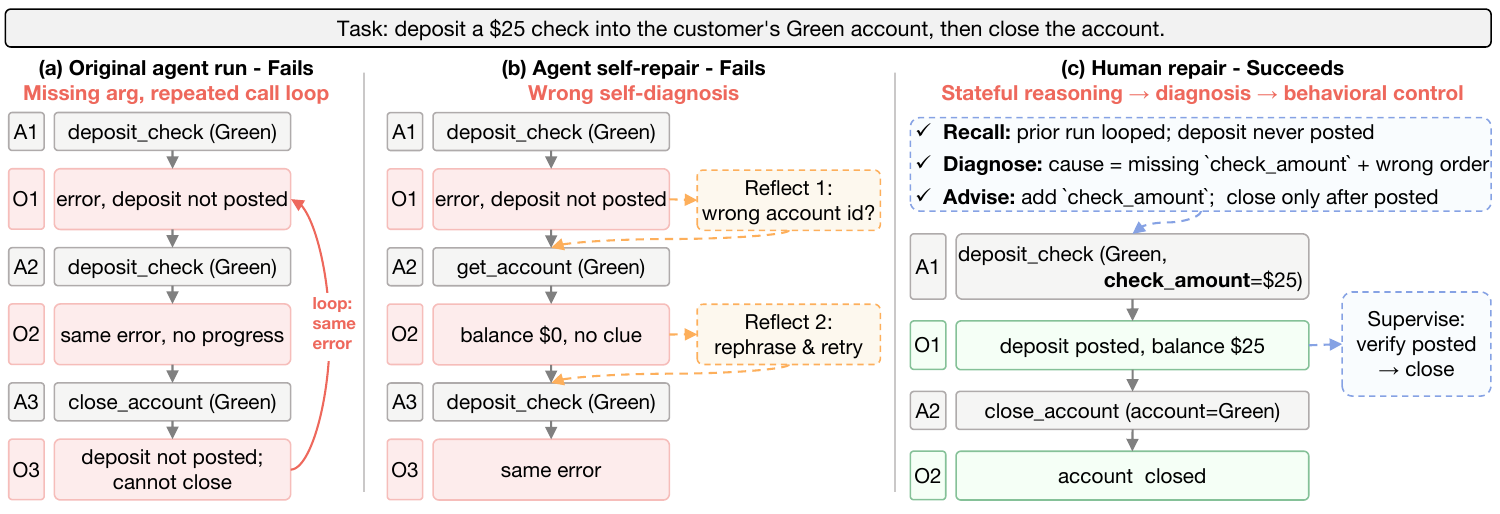}
\caption{Agent self-repair versus human repair on a Banking task. (a) Missing \texttt{check\_amount} prevents the deposit, causing \texttt{close\_account} to fail. (b) Self-repair misdiagnoses the failure and repeats the malformed call. (c) Human repair recalls the prior loop, diagnoses the missing argument, injects corrected guidance, and supervises closure after the deposit is posted.}
\label{fig:repair-gap}
\end{figure*}

Figure~\ref{fig:repair-gap} illustrates the repair gap on a Banking task that requires posting a \$25 check before closing the customer's Green account.
\texttt{deposit\_check} fails because the agent omits \texttt{check\_amount}, so the deposit is never posted and \texttt{close\_account} remains infeasible.
The original run loops on the malformed call and self-repair misdiagnoses it as an account-information problem (columns a–b), while a human operator recalls the prior loop, injects the missing argument, and supervises closure only after the deposit posts (column c).
Together, diagnosis, behavioral control, and stateful reasoning form an integrated repair loop that current agent self-repair methods often lack.

Existing methods reproduce only part of this loop.
Blind retry adds no diagnosis; outcome feedback reports \emph{whether} a run failed, not \emph{where} or \emph{why}~\cite{lightman2023verify}; self-reflection is frequently vague or self-confirming without external evidence~\cite{shinn2023reflexion,madaan2023selfrefine}; and process-level analyses typically stop at detection rather than closing the loop from diagnosis to run-time repair~\cite{zhang2025whichagent,cemri2025mast}.
Consequently, the same failures recur at run-time, where agents repeat a mistake, fix one issue while regressing on another, drift mid-execution, or over-correct after the problem has already been resolved.
This shortfall is fundamental rather than incidental. 
Even when a failure is localized with reasonable accuracy, only a small fraction of runs are actually repaired~\cite{zhao2026probe}, because diagnosis alone does not drive a repair.

A diagnosis drives a repair only when it pinpoints the specific decision that went wrong rather than merely the kind of failure, and only when it stays in force as the agent re-executes rather than being consulted once and forgotten.
Meeting these two requirements automatically raises three challenges.
\begin{itemize}
    \item \textbf{C1: Causal localization from correlated traces.}
A trajectory exposes behavioral correlations rather than the root-cause step itself. 
An early wrong argument or missed signal can propagate through many later steps, with long-range and many-to-many dependencies. 
Therefore, root-cause analysis is a temporal--structural attribution problem rather than a flat reading of the traces.
    \item \textbf{C2: Stateful, behavior-scoped guidance.}
There is a gap between knowing why a run failed and making the next run behave correctly.
A diagnosis can be too broad to act on, too specific to generalize, or forgotten once the agent enters a long tool-use trajectory.
Moreover, one retry may fix the original error while regressing on another constraint, so repair requires remembering what has been fixed and what remains unresolved across iterations.
    \item \textbf{C3: Intervening without over-control.}
Even good guidance can decay during re-execution as the agent drifts, repeats old actions, or takes premature state-changing steps.
However, intervening too often can itself derail the task, block legitimate exploration, or cause the agent to over-correct after the original issue is already resolved.
The challenge is therefore deciding when and how strongly to intervene, and when to stop.
\end{itemize}

We propose \name{}, a run-time repair framework that automates the human repair loop through three coordinated components.
(i)~To address C1, \name{} performs graph-guided RCA for \emph{diagnosis}.
We use a graph representation to preserve both temporal order and non-local dependencies among decisions, tool calls, and environment feedback.
\name{} forms a \emph{Critical Transition Graph (CTG)}, where each \emph{Transition Unit}, a decision--execution--feedback cycle, is a node and information-flow dependencies between units are edges.
This lets attribution operate at the level of behavioral decisions and trace a failure backward along dependency paths rather than stopping at temporally adjacent steps.
Over the CTG, an offline HGT detector and a real-time detector localize anomalous structures.
An analyst LLM then states the root cause, turning point, and recovery hints, requiring no labeled failures.
(ii)~To address C2, \name{} produces stateful repair guidance, pairing \emph{behavioral control} with \emph{stateful reasoning}.
A feedback builder converts the analyst's diagnosis and outcome-failure evidence into scoped guidance, including correction directives and an injection plan.
Cross-iteration \emph{Repair Memory} records what was fixed and what remains unresolved, so later iterations preserve useful fixes while preventing regressions.
(iii)~To address C3, \name{} applies guarded run-time intervention, the online side of \emph{behavioral control}.
A harness checks behavior at tool-return and text-response boundaries, detecting loop repetition, expectation deviation, intent drift, and structural checks.
It then injects corrections by appending to a tool result or inserting a synthetic user message, while evidence grounding, cooldowns, and minimal-intervention guards avoid over-control.

We evaluate \name{} on \taubench{}~\cite{yue2024tau}, a widely used benchmark for realistic, policy-governed tool-agent-user interaction.
Unlike outcome-only benchmarks, \taubench{} checks process-level correctness in long, stateful tool-use runs, making it suitable for end-to-end repair evaluation beyond final-response plausibility.
To cover variation in service workflows and agent backbones, we run all three domains, Retail, Airline, and Banking, covering 261 tasks with Qwen3.7-max~\cite{qwen2026qwen37}, and cross-validate Banking with GPT-5.4~\cite{openai2026gpt54}.
\name{} repairs a substantial fraction of initially failed tasks, with the largest gain on the most constrained Banking domain.
Layered ablations show that graph-guided post-run guidance and guarded intervention provide complementary improvements while reducing agent turns and end-to-end method tokens.

This paper makes the following contributions:
\begin{itemize}
\item \textbf{A unified agent repair framework.}
We present \name{}, a run-time repair framework that combines diagnosis, behavioral control, and stateful reasoning in a single automated pipeline, usable either as a post-run diagnostic-and-guidance tool or with online intervention.

\item \textbf{Graph-guided diagnosis.}
We design a graph-guided RCA method built on the \emph{Critical Transition Graph (CTG)}, which localizes failure-critical subtrajectories by fusing an offline normal-behavior model with a real-time graph detector, attributes failures along dependency paths rather than linear adjacency, and requires no labeled failures.

\item \textbf{Stateful and guarded run-time intervention.}
We design a repair mechanism that turns diagnosis into concise, behavior-scoped guidance, carries cross-iteration \emph{Repair Memory} to prevent regressions, and supervises re-execution through a guarded run-time intervention harness that injects corrections while suppressing unnecessary or risky ones.

\item \textbf{Evaluation across domains and models.}
Against \textit{Blind retry}, \textit{Outcome feedback}~\cite{chen2024selfdebug}, and \textit{Reflexion}~\cite{shinn2023reflexion}, \name{} repairs 69.11\% of initially failed Qwen3.7-max tasks, improving over blind retry by 26.02\pp{} overall and 32.53\pp{} on Banking.
A GPT-5.4 Banking study shows the same trend, and guarded run-time intervention adds $+$12.05\pp{} on Qwen3.7-max Banking where compliance decay is severe.
We release \name{} as an installable package with our experimental setup.\footnote{\url{https://anonymous.4open.science/r/AgentTether-9416/}}

\end{itemize}

\section{Motivation}
\label{sec:background}

To understand why diagnosing a failed run is not enough to repair it, we analyze the initially-failed runs on \taubench{}.
This analysis surfaces two problems, each motivating one core component of \name{}: the diagnosis must pinpoint \emph{where} the run went wrong, and it must stay in force as the rerun unfolds.
We quantify the contribution of each mechanism in Sections~\ref{sec:rq1} and~\ref{sec:rq3}.

\begin{figure}[t]
\centering
\includegraphics[width=0.8\columnwidth]{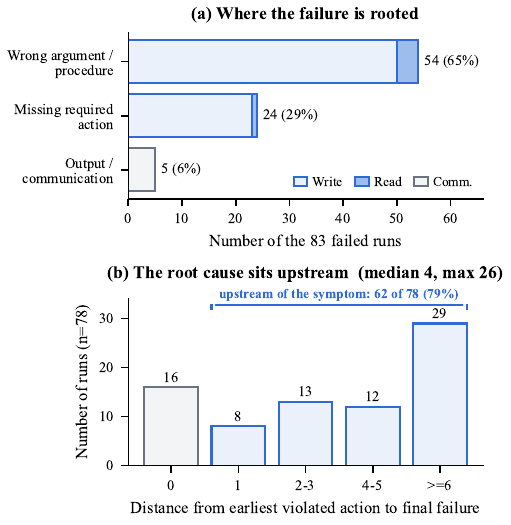}
\caption{Failure analysis of the 83 initially-failed Banking runs (Qwen3.7-max), from \taubench{}'s gold action checks. (a)~Root-cause taxonomy by violated action type. (b)~Causal distance between the earliest violated gold action and the final failure, over the 78 runs with a violated required step.}
\label{fig:failure-taxonomy}
\end{figure}

\textbf{Q1: Can the root cause be localized from a flat trace, or does it require dependency structure?}
We classify all 83 initially-failed Banking runs under Qwen3.7-max by root cause using \taubench{}'s gold action checks, locating for each run the \emph{earliest} required step the agent violated.
Failures are overwhelmingly behavioral: 78 of 83 (94\%) stem from a wrong or missing tool action (65\% a wrong argument or procedure, 29\% a missing required action), and only 6\% are pure communication errors.

A natural baseline is to hand the entire trace to an LLM and ask it to name the root cause. 
This is unreliable for two structural reasons, neither of which a simple positional or frequency heuristic can fix.
\emph{(i)~The cause is far upstream of the symptom, so recency fails.}
The failure signal (\texttt{reward=0}) surfaces only at the end, yet the responsible step lies well before it.
For each of the 78 behavioral failures (the 5 communication-only failures violate no required step), we measure the distance from the earliest violated gold action to the final failure.
The root cause precedes the symptom by a median of 4 required steps (up to 26) and is strictly upstream in 62 of 78 runs (79\%; Figure~\ref{fig:failure-taxonomy}b).
A flat reading, like a recency heuristic, anchors on the last or most salient error, which is merely \emph{correlated} with the failure; the true cause is too far upstream to be picked out by position alone.
\emph{(ii)~Effects are many-to-many, so frequency fails.}
A single wrong decision propagates. 
One root error is followed by 3.2 further violated checks on average, and 76\% of the 78 failures violate more than one check.
A heuristic that flags the most error-dense region therefore cannot separate the originating step from the many downstream steps that merely inherit its effect.
Localizing the cause requires more than reading or counting the trace. 
It requires knowing \emph{which} earlier step's output each later decision consumed, namely the information-flow and dependency edges of the execution.
Modeling these as a graph lets attribution follow causal paths to the originating step, rather than the linear adjacency a flat trace exposes.

\begin{findingbox}
\textbf{Finding~1 (Challenge~1).} Localize root causes over a trajectory graph that models information-flow dependencies, rather than over the final outcome or a flat trace handed to an LLM.
\end{findingbox}

\begin{figure}[t]
\centering
\includegraphics[width=0.8\columnwidth]{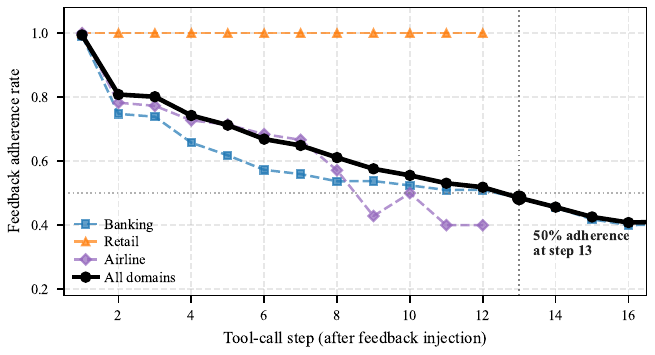}
\caption{Feedback adherence by tool-call step after feedback injection ($n{=}173$ post-feedback reruns, across all three domains). Adherence decays sharply, with Banking the first domain to fall below 50\%.}
\label{fig:compliance-decay}
\end{figure}

\textbf{Q2: Does a correct diagnosis stay in force as the rerun unfolds?}
A precise diagnosis is delivered as a one-shot message injected at the start of the rerun, where it must survive an execution that often runs for more than 20 tool-call steps, alongside task instructions, tool outputs, and domain policy.
To test whether it does, we measure feedback adherence over the 173 post-feedback reruns of a diagnosis-only setting (\textit{\name{} (post-run only)}) approach, detailed in Section~\ref{sec:eval}; iterations~$\geq$1, all three domains).
From each injected report we parse the concrete directives it imposes, namely the tools or actions it forbids (``do not call \texttt{X}'' and the report's boundary constraints), the required actions it flags as missing, and the prohibited behaviors such as shell-based policy lookups.
Walking the rerun's tool calls in order, a run is \emph{adherent} at step~$k$ if it has not yet violated any of these directives; once it violates one, it is counted as fallen for all later steps, yielding a first-violation survival curve.

Figure~\ref{fig:compliance-decay} plots the fraction of reruns still adherent at each step.
Adherence starts at 99\% immediately after injection but, outside the short Retail tasks that need few tool calls, erodes as runs progressively break the feedback. 
It falls to 71\% by step~5 and crosses 50\% at step~13 overall, as early as step~8 on Banking, the longest and most constraint-heavy domain.
In other words, the agent honors a correct diagnosis at first, then drifts back toward its prior behavior, repeats unproductive tool calls, or over-executes after the fix is already in place.
Because such deviations surface only after the feedback has scrolled out of the agent's effective attention window, no improvement to the post-hoc report alone can prevent them.

\begin{findingbox}
\textbf{Finding~2 (Challenges~2--3).} A one-shot diagnosis is not enough; the guidance should be carried across the rerun and applied as it unfolds, catching deviations as they occur and stopping once the repair is complete.
\end{findingbox}

These two problems, the lack of dependency-aware localization and the run-time decay of an otherwise-correct diagnosis, motivate the two core mechanisms of \name{}, namely graph-based diagnosis and guarded mid-execution intervention.

\begin{figure*}[t]
\centering
\includegraphics[width=\textwidth]{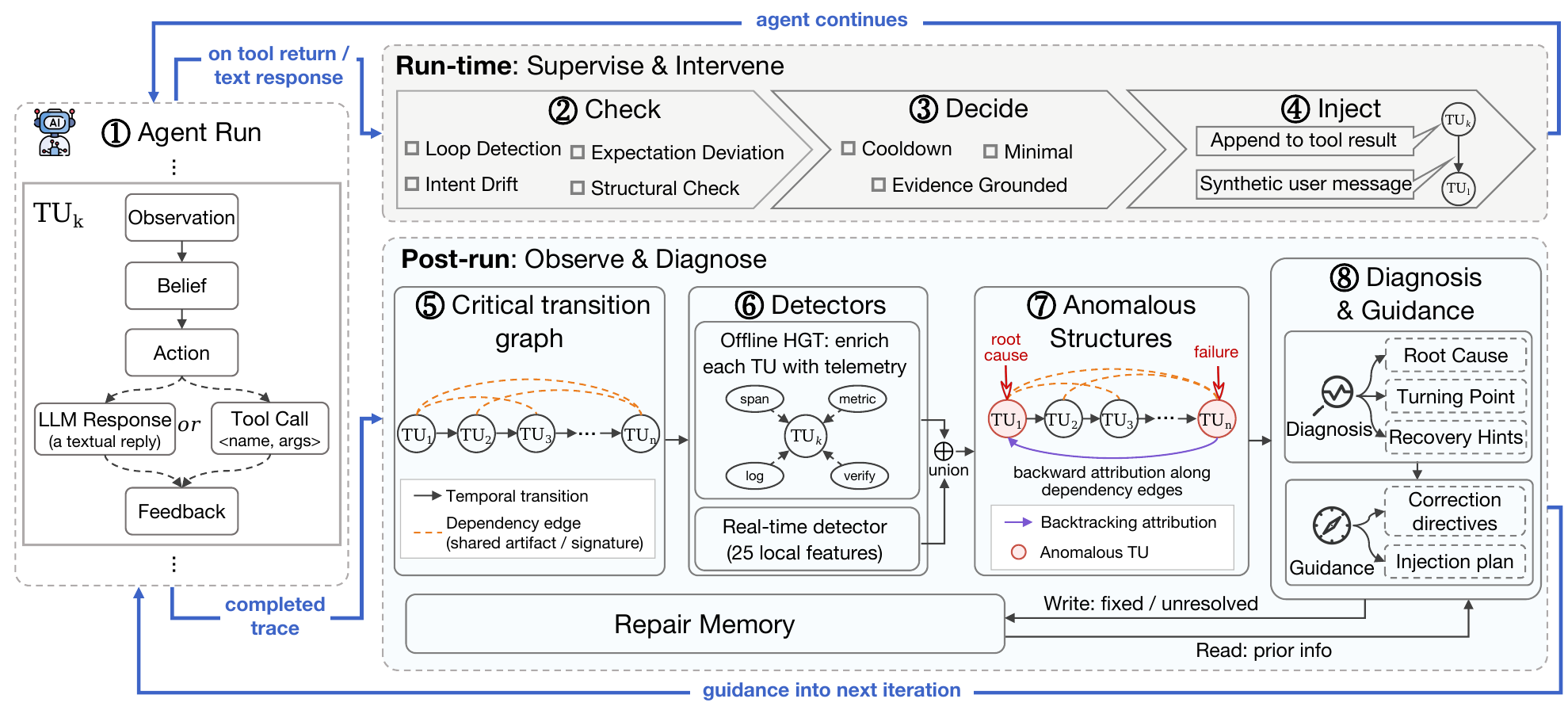}
\caption{Overview of \name{}. 
Two repair tracks wrap a single \emph{Agent Run} and interact only through it. 
\textbf{Post-run (Observe \& Diagnose):} the completed trace is lifted into a Critical Transition Graph (CTG), where two complementary detectors isolate anomalous substructures that an LLM analyst turns into a diagnosis and repair guidance; a cross-iteration \emph{Repair Memory} carries fixed/unresolved state across iterations. 
\textbf{Run-time (Supervise \& Intervene):} a Check $\rightarrow$ Decide $\rightarrow$ Inject harness supervises the next iteration under over-correction guards.}
\label{fig:architecture}
\end{figure*}

\section{The \name{} Approach}
\label{sec:approach}

\subsection{Overview}



As shown in Fig.~\ref{fig:architecture}, \name{} wraps a single unmodified \emph{Agent Run} with two repair tracks that interact with the agent only through its completed trace and subsequent execution.
Each run is represented as an ordered sequence of \emph{Transition Units} (TUs), each capturing one Observation--Belief--Action--Feedback cycle.
The initial run is only observed; \name{} enters the repair cycle only if that run fails.

Between iterations, the \emph{Post-run (Observe \& Diagnose)} track lifts the completed trace into a Critical Transition Graph (CTG), uses the offline HGT detector and real-time detector to localize anomalous substructures, and asks an analyst LLM to turn them into a diagnosis and repair guidance.
This guidance is injected at the start of the next run.
Optionally, the \emph{Run-time (Supervise \& Intervene)} track supervises that re-execution at tool-return and text-response hooks, following a Check--Decide--Inject pipeline to apply guarded corrections when the agent drifts.
The completed run then becomes the next trace for post-run diagnosis, closing the loop.
The cycle stops once a run resolves the task or a fixed budget of $\Gamma$ repair iterations is exhausted.
Cross-iteration \emph{Repair Memory} carries fixed/unresolved state across attempts, so guidance accumulates rather than resets.

\subsection{Agent Execution and Transition Units}

\subsubsection{Trace acquisition}
Diagnosis requires the complete behavioral trace of a run, collected without modifying the agent.
Inspired by distributed tracing~\cite{sigelman2010dapper}, \name{} acquires the trace by monkey-patching known LLM-SDK entry points (e.g., \textit{chat.completions.create}, \textit{messages.create}) with transparent before/after hooks.
Each intercepted event (an LLM inference, a tool invocation, or an environment response) is recorded with its full request/response payload and session context, and appended in execution order to the run's trace.
Patches are removed at session end, leaving the agent's source and behavior unchanged.

\subsubsection{Transition Units}
\name{} abstracts this trace into \emph{Transition Units} (TUs), reasoning at the granularity of \emph{decisions} rather than raw events, and models a run as an ordered sequence $T=\langle \mathrm{TU}_1,\dots,\mathrm{TU}_n\rangle$.
Each $\mathrm{TU}_k=(o_k, b_k, a_k, f_k)$ bundles one decision--execution--feedback cycle: an \emph{observation} $o_k$ (the context the agent conditions on), a \emph{belief} $b_k$ (its natural-language reasoning), an \emph{action} $a_k$ (a tool call $\langle\mathit{name},\mathit{args}\rangle$ or an LLM response), and \emph{feedback} $f_k$ (the execution result and environment output).
Each TU also carries derived attributes that summarize its own outcome and role, including an execution \textit{status} and \textit{error\_signature}, the acting \textit{tool} and its inferred \textit{intent}, and the \textit{artifacts} it produces (e.g., an identifier or a retrieved record).
Beyond these scalar attributes, each TU links to the surrounding run-time telemetry (spans, metrics, logs, and verification records), which the offline detector later represents as typed \emph{context nodes}, grounding diagnosis in signals beyond the agent's own messages.
Working at TU granularity compresses the raw event stream into $O(n)$ behavioral steps and gives every step a stable identity to which evidence (in diagnosis) and corrections (at run-time) attach.

\subsection{Post-run: Observe and Diagnose}

Given the completed TU sequence of a failed run, the post-run track pinpoints \emph{which decision caused the failure, and why}, a temporal--structural attribution problem.
It proceeds in three stages: it builds the trajectory graph, localizes anomalous substructures over it, and analyzes them into repair guidance.

\subsubsection{Trajectory graph}

To model causal relationships within a run, we construct a \emph{Critical Transition Graph} (CTG), denoted as $G=(V,E)$, where nodes $V=\{\mathrm{TU}_1,\dots,\mathrm{TU}_n\}$ correspond to transition units in the execution trace. 
\emph{Temporal} edges $\mathrm{TU}_k \rightarrow \mathrm{TU}_{k+1}$ connect consecutive units and encode execution order.
\emph{Dependency} edges $\{\mathrm{TU}_k, \mathrm{TU}_l\}$ link units that share underlying state (e.g., the same artifact or error signature), enabling non-local attribution beyond temporal adjacency.
Following these edges lets graph-guided RCA trace a failure backward along dependency paths rather than relying only on temporal adjacency.

\subsubsection{Dual anomaly detection}
To localize the failure-critical region on the CTG without any failure labels, two complementary detectors score the trajectory $T$, combining a learned prior of normal behavior with a training-free local detector.

The \emph{offline detector} supplies a learned prior over normal TU behavior.
It operates on a telemetry-enriched view of the CTG, where each TU is augmented with its surrounding typed context nodes, reasoning over these TU-centered neighborhoods to emit a score for each TU.
We train this view on an external success-only corpus of 21,143 resolved trajectories, dominated by TerminalBench~\cite{merrill2026terminalbench} (11,079) and SWE-smith~\cite{yang2025swesmith} (10,010), plus a small number of operational-agent traces.
These sources provide tool-rich, multi-step telemetry and exclude \taubench{} domains and historical runs.
Because these sources differ from \taubench{} in task semantics, we map them into a behavior-oriented telemetry schema that preserves transferable execution structure, including TU order, tool-call/return roles, status/outcome evidence, artifact links, and temporal/dependency edges, while normalizing framework-specific logs and payload layouts.
Thus \name{} learns a normal-execution prior over telemetry structure rather than specific tool policies.
On this corpus, we train a heterogeneous graph transformer~\cite{hu2020heterogeneous} backbone $H$ as a normal-execution model, self-supervised and without any failure or benchmark labels.
It learns normal TU context in two ways: it reconstructs telemetry edges and recovers masked categorical fields (tool, status, role, phase, outcome) from the surrounding graph.

At inference, for each $\mathrm{TU}_k$, the average of its edge-reconstruction and masked-attribute surprises forms a \emph{subgraph-surprise} score $s_{\mathrm{H}}(\mathrm{TU}_k)$, capturing how poorly the local context around $\mathrm{TU}_k$ conforms to the normal model.
\name{} ranks TUs by $s_{\mathrm{H}}$, cuts the anomaly frontier at the largest score drop in the run's own distribution, and groups frontier TUs contiguous in execution order into \emph{frontier groups}.
These groups project back onto the CTG as compact, TU-aligned anomalous subtrajectories.

The \emph{real-time detector} complements the offline HGT detector with run-local evidence.
We instantiate this lightweight run-local detector with Isolation Forest~\cite{liu2012isolation} because it requires no pre-training, works directly on fixed-dimensional TU features extracted from a single CTG, and returns a ranked anomaly score rather than a binary decision.
Given the CTG, the detector maps each TU to a fixed $25$-dimensional graph-contextual feature vector $\phi(\mathrm{TU}_k)$ and obtains a run-local anomaly score $s_{\mathrm{O}}(\mathrm{TU}_k)$.
The features cover three families: \emph{structural} (e.g., in/out degree), \emph{attribute-level} (e.g., status--feedback inconsistency), and \emph{macro-level}, where a \emph{macro} is a contiguous block of TUs that share a sub-goal or tool family (e.g., per-macro error ratio).
High-scoring TUs are expanded along CTG connectivity into localized frontiers.

\name{} keeps the offline and real-time outputs as separate evidence planes: the offline HGT detector captures deviations from successful telemetry structure, while the real-time detector captures outliers within the current CTG.
Their union forms the anomalous substructures $S$.


\subsubsection{Analysis and repair guidance}

The anomalous substructures $S$ localize suspicious regions, but they may still contain redundant detector evidence and routine steps adjacent to those regions.
\name{} therefore prunes $S$ into a compact evidence packet before passing it to the analyst LLM.
For each suspicious region, the packet keeps the peak-scoring TU as a candidate \emph{turning point}, i.e., the transition where the run first leaves a recoverable path even if the failure surfaces later.
It also retains the detector signals supporting this candidate and the minimal local CTG context, including adjacent or dependency-linked TUs, while dropping low-scoring connectors and routine actions.
Given this packet and the run's outcome-failure evidence, the analyst reasons backward along dependency edges and returns a \emph{diagnosis} with three fields: the \emph{root cause}, the confirmed \emph{turning point} $\mathrm{TU}_{t^\star}$, and \emph{recovery hints} for returning the next run to a correct path.

The diagnosis explains why the previous run failed, but it is retrospective and does not specify how to steer the next run.
A feedback builder converts it into \emph{guidance}: short natural-language instructions inserted into the agent's context.
The guidance pairs \emph{correction directives}, which describe how the behavior around $\mathrm{TU}_{t^\star}$ should change, with an \emph{injection plan}, which specifies where the guidance enters the next iteration.
Three design choices shape the guidance.
First, \emph{disclosure control} keeps directives general rather than instance-specific.
A directive names the failing action family and constraint class, such as ``use the savings-eligible source account'', but avoids hard-coding exact parameter values, so the next run follows the failure pattern rather than a fixed answer.
Second, \emph{prominence} places guidance at the start of the next run's context so the correction is seen early.
Third, \emph{concision} merges overlapping recovery hints and keeps the directives most tied to $\mathrm{TU}_{t^\star}$.

Beyond a single iteration, guidance is stateful through \emph{Repair Memory} $M$.
Before producing new guidance, \name{} reads $M$ to recover which directives were fixed and which remain unresolved; after the next run, it writes the updated state back.
This turns post-run feedback from independent one-shot messages into a repair trajectory that reinforces persistent issues while avoiding regressions on fixes that already succeeded.

\subsection{Run-time: Supervise and Intervene}
\label{sec:intervention}
\name{} implements the run-time track as a lightweight intervention harness around the agent execution, without modifying the agent's internal policy.
To keep the re-execution aligned with the active intent, formed by the original task goal and the post-run repair guidance, the harness monitors the agent's behavior and intervenes only when the current trajectory begins to drift.
As shown in Fig.~\ref{fig:architecture}, the harness follows a \textbf{Check\,$\rightarrow$\,Decide\,$\rightarrow$\,Inject} pipeline at two hooks: \emph{on tool return}, after a tool executes but before its result is passed back to the agent, and \emph{on text response}, after the agent generates text but before the response is accepted.
The first hook steers the next decision after observing a tool outcome, while the second intercepts premature or misaligned textual continuations.

\subsubsection{Check}
At each trigger point, the harness evaluates the current action against the active intent and recent execution history through four complementary signals.
Together, they cover the run-time drifts highlighted in Fig.~\ref{fig:architecture}: loop repetition, intent drift, expectation deviation, and delayed response to unresolved guidance.

\emph{Loop detection} catches unproductive repetition by flagging repeated calls to the same tool with identical arguments.
\emph{Intent drift} embeds the active intent and current action, scores their semantic dissimilarity, and applies risk-tiered thresholds: destructive actions are treated most sensitively, information-gathering actions least sensitively, and high-risk drift candidates are confirmed by an LLM verifier.
At tool-return checkpoints, an exponential moving average over recent scores prevents a single exploratory step from triggering intervention.
\emph{Expectation deviation} checks whether expected corrective actions from the guidance or \emph{Repair Memory} have been satisfied, using observed tool names and selected arguments.
If the agent attempts an unexpected destructive operation, the harness raises a warning that names the violated action family or constraint without exposing exact expected parameter values.
\emph{Structural checks} handle delayed or forgotten guidance.
After sufficient progress, or before a submission-like operation, the harness checks whether expected corrective actions remain unsatisfied; if so, it extracts the most relevant correction hint from the active guidance and emits a sparse checkpoint reminder.

\subsubsection{Decide}
Rather than acting on every candidate deviation from \emph{Check}, the \emph{Decide} step applies three guards before intervening.
First, to act only on well-supported problems, \emph{Evidence-grounded} requires each intervention to be backed by an explicit \emph{Check} signal and by active guidance or \emph{Repair Memory}, prioritizing direct evidence such as loops, unexpected destructive actions, or LLM-verified intent drift.
Second, to avoid over-correction, \emph{Cooldown} prevents repeated firings.
It enforces spacing between interventions, caps reroutes on text responses, and deduplicates structural reminders to avoid reissuing the same correction.
Third, to disturb the agent as little as possible,  the \emph{Minimal} guard chooses the least disruptive response.
Once all expected corrective actions are complete, \name{} stops intervening; after an expected state-changing action, it suppresses all non-loop interventions and only intervenes to break repeated calls.

\subsubsection{Inject}
For candidates that pass the \emph{Decide} guards, the \emph{Inject} step delivers a correction through one of two channels.
The correction is generated from the active guidance and the verifier's explanation.
For \emph{on tool return} triggers, the harness appends the correction to the tool result before it is passed back, keeping the intervention in-band with the next decision.
For \emph{on text response} triggers, it injects a synthetic user message and reroutes control to the agent, a stronger channel reserved for text that would otherwise drift or end the run prematurely.
The injected text clarifies the repair direction rather than providing a complete solution.
After injection, the agent continues execution and produces subsequent TUs.

\section{Evaluation}
\label{sec:eval}

Our evaluation addresses three research questions:
\begin{itemize}
\item \textbf{RQ1:} What is the end-to-end repair effectiveness of \name{}, and what does each component contribute?
\item \textbf{RQ2:} Does \name{} transfer across agent model families, and how does repair progress across iterations?
\item \textbf{RQ3:} What is the effect of guarded run-time intervention?
\end{itemize}

\begin{table*}[t]
\centering
\caption{End-to-end performance on initially-failed tasks with Qwen3.7-max. \textbf{Resolved}: tasks repaired / total. \textbf{$\Delta$Retry}: lift over \textit{Blind retry}. \textbf{Avg Steps}: mean agent LLM turns per task, summed over repair iterations. \textbf{Avg Dur.}: mean simulation wall-clock time in minutes. \textbf{E2E Tokens}: average end-to-end method tokens (K) per task, including \name{}'s analyst/verifier calls and excluding benchmark user-simulator/evaluator calls. Overall averages are weighted by initially-failed tasks per domain.}
\label{tab:main-results}
\begin{tabular}{ll cccccc}
\toprule
\textbf{Domain} & \textbf{Approach} & \textbf{Resolved} & \textbf{Rate (\%)} & \textbf{$\Delta$Retry (pp)} & \textbf{Avg Steps} & \textbf{Avg Dur.\ (min)} & \textbf{E2E Tokens (K)} \\
\midrule
\multirow{5}{*}{Retail ($n{=}26$)}
  & Blind retry    & 23/26          & 88.46          & 0.00               & 31.50    & 5.55   & 102  \\
  & Outcome feedback~\cite{chen2024selfdebug}  & 22/26          & 84.62          & $-$3.85            & 25.92    & 4.75   & 83  \\
  & Reflexion~\cite{shinn2023reflexion}      & 23/26          & 88.46          & $+$0.00            & 24.50    & 4.60   & 79  \\
  & \name{} (post-run only) & 24/26          & 92.31          & $+$3.85            & 20.50    & 4.18   & 68  \\
  & \cellcolor{gray!15}\name{} & \cellcolor{gray!15}\textbf{25/26} & \cellcolor{gray!15}\textbf{96.15} & \cellcolor{gray!15}$+$\textbf{7.69} & \cellcolor{gray!15}18.08 & \cellcolor{gray!15}4.09 & \cellcolor{gray!15}59 \\
\midrule
\multirow{5}{*}{Airline ($n{=}14$)}
  & Blind retry    & 8/14           & 57.14          & 0.00               & 32.93    & 6.39   & 114  \\
  & Outcome feedback~\cite{chen2024selfdebug}  & 9/14           & 64.29          & $+$7.14            & 28.86    & 5.60   & 101  \\
  & Reflexion~\cite{shinn2023reflexion}      & 10/14          & 71.43          & $+$14.29           & 28.64    & 6.17   & 100  \\
  & \name{} (post-run only) & \textbf{11/14} & \textbf{78.57} & $+$\textbf{21.43}  & 25.21    & 4.64   & 94  \\
  & \cellcolor{gray!15}\name{} & \cellcolor{gray!15}\textbf{11/14} & \cellcolor{gray!15}\textbf{78.57} & \cellcolor{gray!15}$+$\textbf{21.43} & \cellcolor{gray!15}27.07 & \cellcolor{gray!15}5.55 & \cellcolor{gray!15}102 \\
\midrule
\multirow{5}{*}{Banking ($n{=}83$)}
  & Blind retry    & 22/83          & 26.51          & 0.00               & 83.16    & 21.38  & 1{,}988  \\
  & Outcome feedback~\cite{chen2024selfdebug}  & 17/83          & 20.48          & $-$6.02            & 84.86    & 17.19  & 2{,}013  \\
  & Reflexion~\cite{shinn2023reflexion}     & 22/83          & 26.51          & $+$0.00            & 81.52    & 19.36  & 1{,}967  \\
  & \name{} (post-run only) & 39/83          & 46.99          & $+$20.48           & 77.17    & 17.88  & 1{,}671  \\
  & \cellcolor{gray!15}\name{} & \cellcolor{gray!15}\textbf{49/83} & \cellcolor{gray!15}\textbf{59.04} & \cellcolor{gray!15}$+$\textbf{32.53} & \cellcolor{gray!15}73.29 & \cellcolor{gray!15}20.88 & \cellcolor{gray!15}1{,}738 \\
\midrule
\multirow{5}{*}{Overall ($n{=}123$)}
  & Blind retry    & 53/123         & 43.09          & 0.00               & 66.52    & 16.32  & 1{,}376  \\
  & Outcome feedback~\cite{chen2024selfdebug}  & 48/123         & 39.02          & $-$4.07            & 66.02    & 13.24  & 1{,}387  \\
  & Reflexion~\cite{shinn2023reflexion}    & 55/123         & 44.72          & $+$1.63            & 63.45    & 14.74  & 1{,}356  \\
  & \name{} (post-run only) & 74/123         & 60.16          & $+$17.07           & 59.28    & 13.47  & 1{,}153  \\
  & \cellcolor{gray!15}\name{} & \cellcolor{gray!15}\textbf{85/123} & \cellcolor{gray!15}\textbf{69.11} & \cellcolor{gray!15}$+$\textbf{26.02} & \cellcolor{gray!15}56.36 & \cellcolor{gray!15}15.59 & \cellcolor{gray!15}1{,}197 \\
\bottomrule
\end{tabular}
\end{table*}

\subsection{Experimental Setup}

\textbf{Benchmark.}
We evaluate on \taubench{}~\cite{yue2024tau}, a benchmark for tool-agent-user interaction featuring realistic multi-step tasks with verifiable database state changes, communication requirements, and natural language assertions.
We use three domains of increasing complexity: \emph{Retail} for e-commerce order management (114~tasks), \emph{Airline} for flight booking operations (50~tasks), and \emph{Banking} for financial transactions with strict compliance requirements (97~tasks), totaling 261 tasks.


\textbf{Models.}
The primary agent under repair uses Qwen3.7-max~\cite{qwen2026qwen37} via Alibaba Cloud Model Studio/DashScope, chosen as a strong contemporary Qwen tool-use model with feasible cost for full-domain evaluation.
We additionally evaluate GPT-5.4~\cite{openai2026gpt54} via the OpenAI API on Banking to test cross-model transfer.
To avoid same-model self-judging, all non-agent auxiliary roles, including the \taubench{} user simulator/evaluator and \name{}'s analyst/verifier, use DeepSeek-V4-Pro~\cite{deepseek2026models}, a separate model family with long-context and tool-calling support.
These auxiliary roles are fixed across approaches, so each ablation is compared under the same simulator, evaluator, analyst, and verifier.
For retrieval embeddings, we follow the \taubench{} AllTools setting and use \texttt{text-embedding-3-large}.

\textbf{Repair protocol and approaches.}
All approaches share the same initial execution.
The experiments operate solely on tasks that initially failed, with a maximum of $\Gamma{=}3$ repair iterations after the initial run, ensuring that the evaluation focuses on tasks that genuinely require repair.
Each approach includes or removes a specific \name{} component to isolate its contribution:
\begin{itemize}
\item \textit{Blind retry}: a blind-retry baseline with none of \name{}'s components; the agent reruns for up to three repair iterations with no feedback between iterations.
\item \textit{Outcome feedback}~\cite{chen2024selfdebug}: removes \name{}'s graph-guided diagnosis and gives the next repair iteration only the evaluator's outcome-level assessment of the previous failed run, following the self-debugging pattern of using external feedback for repair.
\item \textit{Reflexion}~\cite{shinn2023reflexion}: replaces \name{}'s external diagnosis with the agent's own self-reflection, generating verbal feedback from its failed trajectory.
\item \textit{\name{} (post-run only)}: \name{}'s post-run track alone, injecting graph-guided diagnosis and repair guidance (backed by Repair Memory) at each retry, with run-time intervention disabled.
\end{itemize}
\name{} adds the guarded run-time intervention on top of \textit{\name{} (post-run only)}.
Unless otherwise noted, all hyperparameters in \name{} are fixed across domains and models and are not tuned per benchmark.
For the offline detector, the HGT normal-execution model uses hidden dimension 52 and is trained for 20 epochs with batch size 128; for the real-time detector, Isolation Forest uses 100 trees, maximum subsample size 256, one-hop graph-context features, and seed 42.
For auxiliary LLM calls, we use low-variance decoding settings where supported, and keep all model parameters fixed across approaches, domains, and agent backbones.
For \emph{Intent drift}, thresholds are 0.35 for destructive actions, 0.50 by default, and 0.55 for information-gathering operations, with five-step exponential moving average smoothing ($\alpha{=}0.3$).
For \emph{Structural checks}, checkpoint reminders require at least five prior steps and five tool calls, allow one repeated reminder, and are capped at three fires per run.

\textbf{Metrics and statistics.}
Our primary metric is task-level binary success rate.
We also report action-level correction rate, which measures whether required tool actions and arguments that were wrong in the initial run are corrected or regress across repair iterations.
To assess efficiency, we report average interaction steps, wall-clock time, and end-to-end method tokens per task.
The token count includes input and output for each agent generation and adds \name{}'s analyst/verifier calls when used.
We exclude benchmark user-simulator/evaluator calls, which are fixed across compared approaches and are not part of the repair method.
Since each task is evaluated under every approach, we use McNemar's exact test on paired binary outcomes.
This test compares how many tasks are resolved by one approach but not by the other.

\subsection{RQ1: End-to-End Repair Effectiveness}
\label{sec:rq1}

\subsubsection{Overall repair effectiveness}
Table~\ref{tab:main-results} reports end-to-end repair results for Qwen3.7-max on all initially failed tasks.
Across the three domains, \name{} repairs 85 of 123 failed runs, corresponding to a 69.11\% repair rate.
This improves over \textit{Blind retry} by 26.02 percentage points and over \textit{AgentTether (post-run only) } by 8.94 points.
The result indicates that both stages of \name{} contribute: graph-guided post-run diagnosis raises the overall repair rate from 43.09\% to 60.16\%, and guarded run-time intervention further raises it to 69.11\%.

The largest domain-level gain appears on Banking, where \name{} repairs 49 of 83 failed runs and improves over blind retry by 32.53 points.
Banking contains long-horizon constraints, compliance checks, and state-changing operations, making failures difficult to recover through repeated execution alone.
The improvement in this domain indicates that \name{} addresses cases in which repair requires identifying an upstream decision and preserving the correction during re-execution.

\subsubsection{Comparison with simpler repair signals}
The baseline approaches separate \name{} from retrying or adding generic feedback.
\textit{Blind retry} repairs 43.09\% overall, showing that some failures can be recovered through stochastic re-execution, but only 26.51\% on Banking.
\textit{Reflexion} repairs 44.72\% overall, only 1.63 points above blind retry, and does not improve Banking at all.
This suggests that the agent's own self-critique often fails to identify the upstream decision that caused the failure.
\textit{Outcome feedback} performs worse than blind retry overall (39.02\%), indicating that final evaluator feedback alone does not provide enough information to guide the next iteration and may direct the agent toward surface symptoms rather than the failure-inducing transition.
By contrast, \textit{\name{} (post-run only)} repairs 60.16\%, showing the value of graph-guided post-run guidance before adding run-time intervention.

\subsubsection{Localization of graph-guided RCA}
We directly test whether graph-guided RCA localizes the upstream failure region from Fig.~\ref{fig:failure-taxonomy}.
Over the 78 behavioral Banking failures under Qwen3.7-max, excluding the 5 communication-only runs, let $g$ be the TU position of the earliest violated \taubench{} gold action and $\hat{p}$ the peak-scoring TU retained from the anomalous substructures $S$ before analyst confirmation.
This proxy is conservative: $g$ marks the first externally visible violation, while graph-guided RCA may select the earlier drift that caused it.
Even under this strict proxy, the anomalous substructures $S$ cover $g$ in 56/78 runs (71.8\%), showing that the graph evidence packet usually contains the failure-critical area.
$\hat{p}$ is also closer to $g$ than position- or density-based shortcuts: median error is 5.5 TUs for \name{}, compared with 6.0 for \emph{Recency} and 10.0 for \emph{Frequency}.
These results indicate that graph attribution is more robust than reading the trace backward from the final symptom or choosing the densest error region.

\subsubsection{Component ablation}
Table~\ref{tab:component-ablation} isolates the contribution of two internal components of \name{}.
To test whether the learned normal-behavior prior adds value beyond run-local anomaly scoring, we remove the offline HGT detector and leave localization to the real-time detector alone.
To test whether cross-iteration state is needed, we remove Repair Memory, preventing the system from recording which corrections have already been fixed and which remain unresolved.

Removing either component reduces repair effectiveness.
Without the offline HGT detector, overall repair falls from 69.11\% to 46.34\%, with the largest drop on Banking (59.04\% to 27.71\%).
Without Repair Memory, overall repair falls to 52.03\%, with Banking dropping to 34.94\%.
These results indicate that the learned normal-execution prior helps identify the right failure region, while Repair Memory helps preserve the correction across multi-iteration repairs.

\begin{table}[t]
\centering
\caption{Component ablation of \name{} on initially-failed tasks (Qwen3.7-max). Values are repair rate (\%); per-domain denominators match Table~\ref{tab:main-results} ($n{=}26/14/83$). Overall is weighted by initially-failed tasks per domain.}
\label{tab:component-ablation}
\begin{tabular}{l cccc}
\toprule
\textbf{Approach} & \textbf{Retail} & \textbf{Airline} & \textbf{Banking} & \textbf{Overall} \\
\midrule
\rowcolor{gray!15}\name{} (full)   & 96.15 & 78.57 & 59.04 & 69.11 \\
\quad w/o offline HGT              & 92.31 & 71.43 & 27.71 & 46.34 \\
\quad w/o Repair Memory            & 92.31 & 78.57 & 34.94  & 52.03    \\
\bottomrule
\end{tabular}
\end{table}

\subsubsection{Interaction budget and latency}
Compared with \textit{Blind retry}, \textit{AgentTether (post-run only) } reduces average agent turns from 66.52 to 59.28 and estimated end-to-end method tokens from 1{,}376K to 1{,}153K, while increasing the repair rate from 43.09\% to 60.16\%.
\name{} uses 56.36 turns and 1{,}197K end-to-end method tokens per task on average, about 13\% fewer tokens than blind retry, while achieving the highest repair rate.
This pattern indicates that trajectory-localized guidance and run-time checkpoints reduce unproductive re-execution instead of simply extending the interaction.
Wall-clock time follows a different pattern because it includes auxiliary verification and intervention latency.
Overall, \name{} is faster than blind retry (15.59 vs.\ 16.32 minutes) but slower than \textit{AgentTether (post-run only) } (15.59 vs.\ 13.47 minutes).
This reflects a trade-off in which guarded run-time intervention improves repair rate and reduces agent-side turns, while introducing additional wall-clock overhead outside the agent's ordinary tool-call sequence.

\subsection{RQ2: Cross-Model Transfer and Repair Dynamics}
\label{sec:rq2}

\begin{figure}[t]
\centering
\includegraphics[width=0.8\columnwidth]{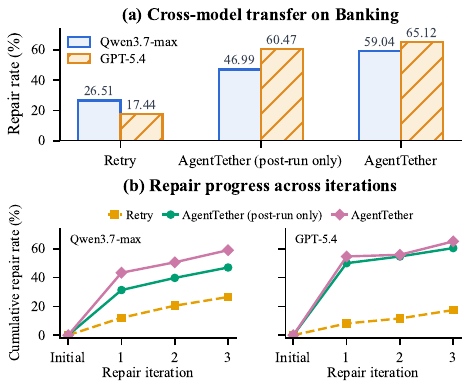}
\caption{Banking transfer and repair dynamics. (a) Cross-model repair rate on initially failed tasks. (b) Cumulative repair rate across iterations, where \emph{Initial} denotes the shared failed run and iterations~1--3 are guided retries. Repair is front-loaded at the first guided iteration for both agents.}
\label{fig:banking-transfer}
\end{figure}

\subsubsection{Cross-model transfer}
Fig.~\ref{fig:banking-transfer}(a) compares Qwen3.7-max with GPT-5.4 on Banking.
We use Banking because it leaves enough initially failed tasks for both agents (83 for Qwen3.7-max and 86 for GPT-5.4), treating these as model-specific repair sets rather than a base-model ranking.
The results indicate that graph-guided diagnosis transfers across model families.
For Qwen3.7-max, \textit{AgentTether (post-run only) } repairs 46.99\% of initially failed tasks, improving over blind retry by 20.48 percentage points; for GPT-5.4, it repairs 60.47\%, improving over blind retry by 43.03 points.
The larger lift on GPT-5.4, mirroring the gain on Qwen3.7-max, shows that graph-guided post-run guidance is not tied to a single model family.
Adding run-time intervention improves both agents: \name{} reaches 59.04\% (49/83) on Qwen3.7-max, adding 12.05 points over \textit{AgentTether (post-run only) }, and 65.12\% (56/86) on GPT-5.4, adding 4.65 points.
The smaller intervention gain on GPT-5.4 reflects fewer residual drift cases after graph-guided feedback alone, whereas Qwen3.7-max benefits more from run-time checks because it more often drifts from the injected repair guidance during re-execution.

\subsubsection{Repair dynamics}
Fig.~\ref{fig:banking-transfer}(b) plots cumulative repair rate on Banking across repair iterations.
Repair is concentrated at iteration~1, the first retry after guidance is introduced.
With \name{}, 43.37\% of initially failed Qwen3.7-max Banking tasks are repaired at iteration~1, accounting for 73.5\% of all eventual repairs.
For GPT-5.4, 54.65\% are repaired at iteration~1, accounting for 83.9\% of eventual repairs.
Later iterations provide substantially smaller marginal gains.
This pattern supports the bounded repair protocol used by \name{}, where a small number of repair iterations captures most recoverable cases.


\subsection{RQ3: Effect of Guarded Run-Time Intervention}
\label{sec:rq3}
To isolate guarded run-time intervention, we compare \name{} with \textit{\name{} (post-run only)} on the same initially failed tasks.
Table~\ref{tab:intervention} reports paired \emph{helped} and \emph{hurt} counts, where a task is helped if only \name{} repairs it and hurt if only \textit{\name{} (post-run only)} repairs it.
Intervention yields a significant net gain on Banking ($+10$, $p{=}0.021$), is positive but not significant on Retail ($+1$, $p{=}1.00$), and is neutral on Airline.
This suggests that run-time intervention is most useful when repair requires preserving a specific correction plan through long, state-changing workflows.

\begin{table}[t]
\centering
\caption{Intervention effect by domain: \name{} vs.\ \textit{AgentTether (post-run only) } on paired initially-failed tasks with Qwen3.7-max. \textbf{Helped}: resolved only by \name{}. \textbf{Hurt}: resolved only by \textit{AgentTether (post-run only) }.}
\label{tab:intervention}
\begin{tabular}{@{}lcccl@{}}
\toprule
Domain & Helped & Hurt & Net & Sig. \\
\midrule
Retail ($n{=}26$)  & 1  & 0 & $+$1  & $p{=}1.00$ \\
Airline ($n{=}14$) & 2  & 2 & $+$0  & $p{=}1.00$ \\
Banking ($n{=}83$) & 13 & 3 & $+$10 & $p{=}0.021^{*}$ \\
\bottomrule
\end{tabular}
\end{table}

To further explain the result, we inspect the run-time intervention logs for the \emph{helped} and \emph{hurt} cases in the Banking domain.
The \emph{helped} tasks receive sparse interventions: across the 13 \emph{helped} Banking tasks, \name{} triggers 11.3 interventions per task on average.
All recorded interventions use the lighter \emph{on tool return} channel, reflecting that Banking deviations usually surface after tool execution as missing evidence, unmet preconditions, or premature state-changing actions.
Most correspond to \emph{Structural Check}, with fewer \emph{Expectation Deviation} warnings.
This indicates that interventions in repaired cases usually act as lightweight checkpoints, keeping the agent aligned with the repair guidance while allowing it to continue the task.
The \emph{hurt} cases show a different pattern.
Across the 3 Banking \emph{hurt} cases, \name{} triggers 29.3 interventions per task on average, with a larger share of \emph{Expectation Deviation} warnings on state-changing calls.
These cases demonstrate that when the same guard repeatedly fires around a required action, the agent may re-plan instead of committing to the correct operation.
Thus, while cooldowns and caps limit over-intervention, balancing timely correction against excessive control remains an open policy-design issue.

\section{Discussion}
\label{sec:discussion}

\textbf{Intervention is useful but not free.}
\name{} separates post-run diagnosis from run-time intervention because identifying the failed decision and keeping the correction active during re-execution are different problems.
This distinction explains our results: intervention helps most on Banking, where long trajectories and state-changing operations make one-shot guidance decay quickly, but is neutral on Airline, where legitimate exploration can resemble drift.
The hurt cases show the main risk: repeated guards around a required action may cause the agent to re-plan instead of committing.
Future systems should therefore adapt intervention strength to domain risk, task phase, and model capability.

\textbf{Repair quality depends on auxiliary judgments.}
\name{} uses an analyst for diagnosis and a verifier for intervention, so it does not eliminate dependence on LLM judgments.
Three properties bound this dependence: diagnosis is grounded in CTG-localized telemetry with a graph-algorithmic fallback; the verifier only gates intervention, not task outcome, and over-correction guards limit a wrong call's disruption; and auxiliary roles stay separate from the repaired agent to reduce self-evaluation leakage.
Reducing this reliance via stronger label-free diagnosis or lighter verifiers is a promising direction.

\textbf{Scope and deployment.}
\name{} is not simply a more expensive retry: it reduces agent turns and end-to-end approach tokens while improving repair, and it also recovers failures in the separate GPT-5.4 Banking study.
The main added cost is auxiliary latency, making deployment a choice between offline post-run repair and full run-time intervention.
Our evaluation on \taubench{} covers process-level failures in long, stateful tool-use tasks, but production systems may have different failure distributions, tool ecosystems, and evaluator reliability.
Extending \name{} to code generation, multi-agent coordination, and production observability remains future work.

\section{Related Work}

\begin{table}[t]
\centering
\caption{Positioning of \name{} relative to prior approaches.
\emph{Diagnostic}: locates where/why a run failed from execution evidence;
\emph{Reasoning}: operates at the step level;
\emph{Online}: intervenes during execution;
\emph{Stateful}: carries repair state across iterations;
\emph{Oracle-free}: needs no ground-truth answers or tests.}
\label{tab:positioning}
\footnotesize
\setlength{\tabcolsep}{3pt}
\begin{tabular}{@{}lccccc@{}}
\toprule
System & Diagnostic & Reasoning & Online & Stateful & Oracle-Free \\
\midrule
Reflexion~\cite{shinn2023reflexion}      & \xmark & \xmark & \xmark & \cmark & \cmark \\
Self-Refine~\cite{madaan2023selfrefine}  & \xmark & \xmark & \xmark & \xmark & \cmark \\
SelfDebug~\cite{chen2024selfdebug}       & \xmark & \cmark & \xmark & \xmark & \xmark \\
Failure Attrib.~\cite{zhang2025whichagent}           & \cmark & \cmark & \xmark & \xmark & \xmark \\
PROBE~\cite{zhao2026probe}               & \cmark & \cmark & \xmark & \xmark & \cmark \\
\textbf{\name{}}                         & \cmark & \cmark & \cmark & \cmark & \cmark \\
\bottomrule
\end{tabular}
\end{table}

Table~\ref{tab:positioning} compares prior work along the capabilities needed for agent repair: diagnosis, behavioral control, and stateful reasoning.
Existing methods cover parts of this loop, but none combine trace-grounded diagnosis, process-level reasoning, online intervention, cross-iteration repair state, and oracle-free operation.

\textbf{Feedback-based agent repair.}
Feedback-based methods reuse failed-run signals across attempts.
Reflexion~\cite{shinn2023reflexion} learns from verbal feedback, Self-Refine~\cite{madaan2023selfrefine} runs generate-then-critique loops, and SelfDebug~\cite{chen2024selfdebug} re-executes against test results; tool-interactive critiquing~\cite{gou2024critic}, iterative computer-control agents~\cite{kim2023language}, and studies of self-correction and inner-monologue planning~\cite{kamoi2024can, huang2022inner, huang2024large} explore related feedback channels.
Offline optimizers such as DSPy~\cite{khattab2024dspy} and ProTeGi~\cite{pryzant2023automatic} instead search prompt templates before execution.
These methods support retry- or prompt-level adaptation, but their feedback is not tied to a specific behavioral transition and does not keep a correction active during re-execution.

\textbf{Process-level analysis and run-time monitoring.}
A second line localizes or checks behavior at the step level but stops short of closing the repair loop.
Automated failure attribution~\cite{zhang2025whichagent} localizes the agent and step responsible for a task failure from execution logs.
Classic software-engineering fault localization~\cite{wong2016survey, jones2005empirical} and dependency-graph root cause analysis in distributed systems~\cite{sole2011survey} inspire our graph-based attribution, though they require substantial adaptation to the semantic, non-deterministic nature of LLM traces.
Run-time monitoring methods, including specification-based monitoring~\cite{bartocci2018specification}, process mining~\cite{vanderaalst2012process}, and LM-emulated sandboxing~\cite{ruan2024identifying}, watch executions as they unfold, but mainly enforce safety or specification constraints rather than closing the loop from diagnosis to repair.

\textbf{Failure-anchored recovery.}
PROBE~\cite{zhao2026probe} grounds post-failure recovery in run telemetry, organizing failed-run evidence into a structured diagnosis of the failure \emph{category} and bounded recovery guidance.
It has high accuracy in fault category identification but low repair rates, indicating that naming fault types is necessary but far from sufficient.
\name{} instead performs graph-guided RCA over the Critical Transition Graph (CTG), turns the analyst LLM's diagnosis into behavior-scoped guidance, carries fixed/unresolved state through Repair Memory, and applies guarded run-time intervention when the agent drifts during repair.

\section{Conclusion}
\label{sec:conclusion}

We presented \name{}, a runtime repair framework that bridges the gap between diagnosing failed LLM-agent trajectories and recovering them in subsequent execution.
\name{} abstracts runs into Transition Units, localizes failure-critical subtrajectories over a Critical Transition Graph, and converts the diagnosis into stateful guidance and guarded run-time intervention, all without modifying the underlying agent or environment.
Across 261 \taubench{} tasks and two model families, \name{} improves repair over blind retry while reducing unproductive agent-side execution.
On Banking, it repairs 59.04\% of initially failed Qwen3.7-max tasks and 65.12\% of initially failed GPT-5.4 tasks.
The results show that effective agent repair requires more than identifying a failure: diagnosis must be carried across iterations and enforced only when execution drifts.
Future work includes adaptive intervention policies, multi-agent repair, and integration with production observability platforms.

\section*{Data Availability}
We have made our source code and datasets publicly available through an anonymous repository
at \url{https://anonymous.4open.science/r/AgentTether-9416/}.


\bibliographystyle{IEEEtran}
\bibliography{references}

\end{document}